\documentclass[aps,prl,onecolumn,showpacs,preprintnumbers]{revtex4}
\usepackage{amsmath}
\usepackage{dcolumn}
\usepackage{bm}
\usepackage{subfigure}
\usepackage{amsfonts}
\usepackage{amssymb}
\usepackage{makeidx}
\usepackage{epsfig}
\usepackage{graphicx}

\begin{document}

\title[]{\textbf{Theory of non-local point transformations - Part 3: Theory
of NLPT-acceleration and the physical origin of\ acceleration effects in
curved space-times}\\
}
\author{Massimo Tessarotto}
\affiliation{Department of Mathematics and Geosciences, University of Trieste, Italy}
\affiliation{Institute of Physics, Faculty of Philosophy and Science, Silesian University
in Opava, Bezru\v{c}ovo n\'{a}m.13, CZ-74601 Opava, Czech Republic}
\author{Claudio Cremaschini}
\affiliation{Institute of Physics, Faculty of Philosophy and Science, Silesian University
in Opava, Bezru\v{c}ovo n\'{a}m.13, CZ-74601 Opava, Czech Republic}
\date{\today }

\begin{abstract}
This paper is motivated by the introduction of a new functional setting of
General Relativity (GR) based on the adoption of suitable group non-local
point transformations (NLPT). Unlike the customary local point
transformatyion usually utilized in GR, these transformations map in each
other intrinsically different curved space-times. In this paper the problem
is posed of determining the tensor transformation laws holding for the $4-$%
acceleration with respect to the group of general NLPT. Basic physical
implications are considered. These concern in particular the identification
of NLPT-acceleration effects, namely the relationship established via
general NLPT between the $4-$accelerations existing in different
curved-space times. As a further application the tensor character of the EM
Faraday tensor.with respect to the NLPT-group is established.
\end{abstract}

\maketitle





\section{1 - Introduction}

Following the program outlined in Ref.\cite{noi1} (Part 1) and Ref.\cite%
{noi2} (Part 2) concerning the formulation of the theory of non-local point
transformations (NLPT) for classical tensorial calculus, in this paper the
problem is posed of\ determining the tensor transformation law for the\ $4-$%
accelerations defined in different curved space-times and to ascertain the
consequent possible physical implications in general relativity (GR) which
are inspired by the related transformation theory.

The issue is in some sense intrinsically built in GR and is intimately
connected with the historical origins of the discipline. In particular it is
closely related with the now one-century old foundations of GR dating back
to Albert Einstein's 1915 article on the field equation bringing his name
\cite{Einstein1915} (see also Refs. \cite%
{Einst,Landau,Wheeler,Wheeler2,Synge}). The solution of the problem in the
context of GR is, in fact, almost trivial when formulated in the customary
functional setting initially chosen by Einstein himself \cite%
{ein-1907,ein-1911}, namely when GR-reference frames (i.e., coordinate
systems) are connected by means of local point transformations (LPT) only
and the transformation laws of the related $4-$accelerations are considered.
However, the issue of how to connect the $4-$accelerations defined in
different curved space-time represents in itself a formidable study case
and, to this date, a theoretical challenge \cite{mash1,mash2} This
conclusion actually seems to unveil an amazing aspect of history science and
particularly of GR, namely that the mainstream literature has effectively
ignored the potential of the problem. In the following we intend to show
that the determination of the $4-$acceleration non-local transformation laws
is actually of foremost importance for its implications in GR and all
relativistic theories.

In particular, we intend to show that such a problem can be conveniently
formulated in the context of the theory of non-local point transformations
(NLPT), and its group of general NLPT $\left\{ P_{g}\right\} ,$ presented
and discussed in detail in Parts 1 and 2.

In fact, although the mathematical adequacy of the classical theory of
tensor calculus on manifolds, lying as the basis of GR, remains paramount,
compelling theoretical motivations suggest the actual \textit{physical
inadequacy} of the traditional setting of LPT adopted both in GR and
relativistic theories. Nevertheless, as shown in Part 1, in certain physical
problems including in particular Einstein's Teleparallel approach to the
theory of gravitation \cite{Einstein1928} (see also Refs \cite%
{tele1,tele2,tele3})and the problem of diagonalization of metric tensors in
GR \cite{NJ1,NJ2,NJ3} (see also Refs.\cite{bambi,bambi2}), the introduction
of the NLPT-setting is found to be mandatory.\textbf{\ }From the
investigation carried out in Parts 1 and 2 it follows that the group of
general NLPT $\left\{ P_{g}\right\} $ can map in each other two in principle
arbitrary connected and time-oriented Lorentzian curved space-times $(%
\mathbf{Q}^{4},g)$ and $(\mathbf{Q}^{\prime 4},g^{\prime })$. In\ Lagrangian
and Eulerian forms these maps are respectively of the type%
\begin{equation}
\left\{
\begin{array}{c}
P_{S}\text{: }r^{\mu }(s)=r^{\mu }(s_{o})+\int_{s_{o}}^{s}d\overline{s}%
M_{(g)\nu }^{\mu }(r^{\prime },r(r^{\prime }))u^{\prime \nu }(\overline{s}),
\\
P_{S}^{-1}\text{: }r^{\prime \mu }(s)=r^{\mu }(s_{o})+\int_{s_{o}}^{s}d%
\overline{s}\left( M_{(g)}^{-1}\right) _{\nu }^{\mu }(r,r^{\prime
}(r))u^{\nu }(\overline{s}),%
\end{array}%
\right.  \label{g-NLPT}
\end{equation}%
and%
\begin{equation}
\left\{
\begin{array}{c}
P_{S}\text{: }r^{\mu }=r_{o}^{\mu }+\int_{r_{o}^{\prime \mu }}^{r^{\prime
\mu }}dr^{\prime \nu }M_{(g)\nu }^{\mu }(r^{\prime },r(r^{\prime })), \\
P_{S}^{-1}\text{: }r^{\prime \mu }=r_{o}^{\prime \mu }+\int_{r_{o}^{\mu
}}^{r^{\mu }}dr^{\nu }\left( M_{(g)}^{-1}\right) _{\nu }^{\mu }(r,r^{\prime
}(r)).%
\end{array}%
\right.  \label{g-NLPT-Eulerian}
\end{equation}%
Here the notations are those pointed out in Part 2. Thus, in particular%
\textbf{\ }$r^{\mu }(s)$ and $r^{\prime \mu }(s)$ denote suitably smooth
world-lines referred to arbitrary curvilinear coordinate systems $r\equiv
\left\{ r^{\mu }(s\right\} )$ and $r^{\prime }\equiv \left\{ r^{\prime \mu
}(s)\right\} $ of $(\mathbf{Q}^{4},g)$ and $(\mathbf{Q}^{\prime 4},g^{\prime
})$ respectively, while $u^{\nu }(\overline{s})=dr^{\mu }/ds\ $and $%
u^{\prime \nu }(\overline{s})=dr^{\prime \mu }/ds$ are the corresponding $4-$%
velocities.\ The corresponding $4-$velocity transformations are then%
\begin{equation}
\left\{
\begin{array}{c}
u^{\mu }(s)=M_{(g)\nu }^{\mu }(r^{\prime },r(r^{\prime }))u^{\prime \nu }(s),
\\
u^{\prime \mu }(s)=\left( M_{(g)}^{-1}\right) _{\nu \nu }^{\mu \mu
}(r,r^{\prime }(r))u^{\nu }(s).%
\end{array}%
\right.  \label{4-velocity trasnformations}
\end{equation}%
Here $M_{(g)\nu }^{\mu }(r^{\prime },r(r^{\prime })),\left(
M_{(g)}^{-1}\right) _{\nu }^{\mu }(r,r^{\prime }(r))$ denote the Jacobian
matrix and its inverse, both to be assumed of \emph{non-gradient type} (see
related definitions in Part 1). In particular, these can be represented
respectively as the two real Jacobian matrices%
\begin{eqnarray}
M_{(g)\nu }^{\mu }(r^{\prime },r^{\prime }) &=&\frac{\partial g_{A}^{\mu
}(r^{\prime })}{\partial r^{\prime \nu }}+A_{(g)\nu }^{\mu }\left( r^{\prime
},\left[ r^{\prime },u^{\prime }\right] \right) ,  \label{poi5} \\
\left( M_{(g)}^{-1}\right) _{\nu }^{\mu }(r,r^{\prime }) &=&\frac{\partial
f_{A}^{\mu }(r)}{\partial r^{\nu }}+B_{(g)\nu }^{\mu }\left( r,\left[ r,u%
\right] \right) ,  \label{poi6}
\end{eqnarray}%
where the functions on the rhs of the previous equations are suitably
defined (see Paper 2).

However, a basic issue that arises in GR, but which includes also classical
and quantum mechanics as well as the theory of classical and quantum fields,
is the role of non-local effects arising due to the choice of the extended
GR-frames, namely of the state vectors%
\begin{eqnarray}
\mathbf{x}(s) &\equiv &\left\{ r^{\mu }(s),u^{\mu }(s)\right\} ,
\label{state-vectors-1} \\
\mathbf{x}^{\prime }(s) &\equiv &\left\{ r^{\prime \mu }(s),u^{\prime \mu
}(s)\right\} ,  \label{state-vectors-2}
\end{eqnarray}%
which are defined at the same prescribed proper time $s$ and for all $s\in
I\subseteq
\mathbb{R}
.$\textbf{\ }The physical motivations are based on the Einstein principle of
equivalence, namely ultimately on the equivalence between accelerating
reference frames and the occurrence of gravitational fields, in connection
with intrinsically different space-times.\textbf{\ }For this purpose in this
paper we intend to focus our attention on the role of acceleration effects
in GR, arising specifically due to NLPT transformations between different
space-times. In fact, the precise mathematical formulation and physical
mechanisms by which non-locality should manifest itself between accelerating
frames must still be fully understood.

\subsection{Goals of the paper}

Following the recent proposal of extending the class of reference frames on
which GR relies, to include the treatment of a suitably-prescribed group $%
\left\{ P_{g}\right\} $ of non-local point transformations, in this paper
the problem is posed of analyzing the mathematical properties of the
corresponding Jacobian matrix and the related tensor transformation laws. As
a consequence, it is shown that in such a setting, namely for arbitrary NLPT
of the group $\left\{ P_{g}\right\} ,$ the relativistic $4-$acceleration is
covariant, i.e., is endowed with $4-$tensor transformation laws.

More precisely, the goals and structure of the paper are as follows.

\begin{enumerate}
\item \emph{GOAL \#1 - }The first one, discussed in\ Section 2, concerns the
analysis of the differential properties of the Jacobian matrix $M_{(g)\nu
}^{\mu }$ associated with the general NLPT-group $\left\{ P_{g}\right\} $
(see Lemmas 1 and 2).

\item \emph{GOAL \#2 - }In Section 3 the transformation properties of the
Christoffel symbols with respect of the same group are displayed (THM.1).
The mapping existing between the Christoffel symbols corresponding to
different curved space-times connected by a general NLPT is displayed. As a
special example the particular case of special NLPT mapping the Minkowski
space-time in a generic curved space-time is considered.

\item \emph{GOAL \#3 - }In Section 4, the $4-$vector transformation
properties of the $4-$acceleration with respect to the general NLPT-group $%
\left\{ P_{g}\right\} $ are determined (THM.2). Remarkably the $4-$%
acceleration are shown to transform with respect to general NLPT as $4-$%
vectors.

\item \emph{GOAL \#4 - }In Section 5, as a first application of THM.2, the
acceleration transformation equations are determined by means of NLTP having
purely diagonal-Jacobian matrices which map Schwarzschild or Reissner-Nordstr%
\"{o}m space-times either onto a flat Minkowski or Schwarzschild-analog
space-times.

\item \emph{GOAL \#5 - }In Section 6, the application concerns the treatment
of the Friedmann-Lemaitre-Robertson-Walker (FLRW) curved space-time. This is
found to be mapped onto the flat Minkowski space-time by means of NLPT
having a non-diagonal Jacobian matrix.

\item \emph{GOAL \#6 - }In Section 7, the application is considered of the
Kerr-Newman space-time, similarly mapped on Schwarzschild-analog space-times.

\item \emph{GOAL \#7 - }In Section 8, the application is considered which
concerns the prescription of the tensor transformation laws of the
electromagnetic (EM) Faraday tensor with respet to the group of general NLPT
$\left\{ P_{g}\right\} $.

\item \emph{GOAL \#8 - }Finally, in Section 9 the main conclusions of the
paper are drawn.
\end{enumerate}

\section{2 - Mathematical preliminaries: differential properties of $M_{(g)%
\protect\nu }^{\protect\mu }$}

In this section the relevant properties of the Jacobian $M_{(g)\nu }^{\mu }$
(see Eq.(\ref{poi5}) and Eq.(\ref{poi6}) for the corresponding inverse
matrix) which is associated with a generic transformation of the group $%
\left\{ P_{g}\right\} $ are summarized.

For the sake of reference, let us consider first the case in which the point
transformations given by Eqs.(\ref{g-NLPT}) reduce to the customary form of
local point transformations. This case occurs manifestly if the matrices $%
A_{(g)\nu }^{\mu }$ and $B_{(g)\nu }^{\mu }$ vanish identically so that the
transformations reduce to the $C^{(k)}-$diffeomorphism (with $k\geq 3$)%
\begin{equation}
r^{\mu }(s)=g_{A}^{\mu }(r^{\prime }(s)),  \label{poi1-L}
\end{equation}%
\begin{equation}
r^{\prime \mu }(s)=f_{A}^{\prime \mu }(r(s)),  \label{poi2-L}
\end{equation}%
while the corresponding Jacobian $M_{\nu }^{\prime \mu }$ becomes a local $%
C^{(k-1)}-$function of the form $M_{\nu }^{\prime \mu }=M_{\nu }^{\prime \mu
}(r^{\prime }).$ Hence, the differential of $M_{\nu }^{\prime \mu }$ takes
the form prescribed by the (Leibnitz) chain rule of differentiation. The
following proposition holds.

\bigskip

\textbf{LEMMA 1 - Differential identity for LPT}

\emph{Given validity of Eqs.(\ref{poi1-L}) and (\ref{poi2-L}) the Jacobian }$%
M_{(g)\nu }^{\mu }=M_{\nu }^{\mu }(r^{\prime })$\emph{\ is identified with
the\ }$C^{(k-1)}-$\emph{function}%
\begin{equation}
M_{\nu }^{\mu }(r^{\prime })=\frac{\partial g_{A}^{\mu }(r^{\prime })}{%
\partial r^{\prime \nu }}  \label{idd-00}
\end{equation}%
\emph{so that the differential of the Jacobian }$M_{\nu }^{\mu }(r^{\prime
}) $\emph{\ reads}%
\begin{equation}
dM_{\nu }^{\mu }(r^{\prime })=dr^{\prime \alpha }\frac{\partial M_{\nu
}^{\mu }(r^{\prime })}{\partial r^{\prime \alpha }},  \label{IDD-01}
\end{equation}%
\emph{where on the rhs }$\frac{\partial M_{\nu }^{\mu }(r^{\prime
},r(r^{\prime }))}{\partial r^{\prime \alpha }}$\emph{\ denotes the partial
derivative with respect to} $r^{\prime \alpha }.$

\bigskip

Next, let us consider the case of an arbitrary NLPT, for which the Jacobian $%
M_{(g)\nu }^{\mu }$ is instead of the form $M_{(g)\nu }^{\mu }(r^{\prime
},r) $ (see Eqs.(\ref{poi5})), where $r\equiv r(r^{\prime })\equiv \left\{
r^{\mu }(r^{\prime })\right\} $ and the implicit (and non-local) dependence
in terms of $r^{\prime }\equiv \left\{ r^{\prime \mu }\right\} $ occurring
via $r\equiv \left\{ r^{\mu }\right\} $ is considered as prescribed via the
NLPT. As an example, in the case of a special NLPT (see Paper 1) it follows
that%
\begin{equation}
M_{(g)\nu }^{\mu }(r^{\prime },r)\equiv M_{(g)\nu }^{\mu }(r^{\prime
},r^{\prime }+\Delta r^{\prime }(s)),
\end{equation}%
where $\Delta r^{\prime \mu }(s)$ $\equiv \Delta r^{\prime \mu }$ takes the
form%
\begin{equation}
\Delta r^{\prime \mu }\equiv \int_{r_{o}^{\prime \mu }}^{r^{\prime \mu
}}dr^{\prime \nu }A_{(g)\nu }^{\mu }(r^{\prime },r^{\prime }+\Delta
r^{\prime }).  \label{IDD-1}
\end{equation}%
As a consequence, invoking in particular Eq.(\ref{g-NLPT-Eulerian}), one
obtains respectively:%
\begin{equation}
\frac{\partial r^{\beta }}{\partial r^{\prime \alpha }}=\frac{\partial }{%
\partial r^{\prime \alpha }}\left[ r^{\prime \beta }+\Delta r^{\prime \beta
}(s)\right] =\delta _{\alpha }^{\beta }+A_{\alpha }^{\beta }(r^{\prime
},r^{\prime }+\Delta r^{\prime })\equiv M_{(g)\alpha }^{\beta }(r^{\prime
},r),  \label{IDD-2}
\end{equation}%
\begin{equation}
\left. \frac{\partial M_{(g)\nu }^{\mu }(r^{\prime },r(r^{\prime }))}{%
\partial r^{\beta }}\right\vert _{r^{\prime }}=\frac{\partial r^{\prime
\beta }}{\partial r^{\alpha }}\left. \frac{\partial M_{(g)\nu }^{\mu
}(\left( r^{\prime }\right) ,r(r^{\prime }))}{\partial r^{\prime \beta }}%
\right\vert _{\left( r^{\prime }\right) }=\left( M^{-1}\right) _{\alpha
}^{\beta }\left. \frac{\partial M_{(g)\nu }^{\mu }(\left( r^{\prime }\right)
,r(r^{\prime }))}{\partial r^{\prime \beta }}\right\vert _{\left( r^{\prime
}\right) },  \label{IDD-3}
\end{equation}%
where we have denoted symbolically $r(r^{\prime })\equiv r^{\prime }+\Delta
r^{\prime }$. As a consequence, the following proposition, analogous to that
warranted by Lemma 1 in the case of LPTs, holds.

\bigskip

\textbf{LEMMA 2 - Differential identity for NLPT}

\emph{Given validity of THM.1\ in Part 2 the differential of the Jacobian }$%
M_{(g)\nu }^{\mu }(r^{\prime },r)=M_{(g)\nu }^{\mu }(r^{\prime },r\left(
r^{\prime }\right) )$\emph{\ reads}%
\begin{equation}
dM_{(g)\nu }^{\mu }(r^{\prime },r(r^{\prime }))=dr^{\prime \alpha }\frac{%
\partial M_{(g)\nu }^{\mu }(r^{\prime },r(r^{\prime }))}{\partial r^{\prime
\alpha }},  \label{IDD-4}
\end{equation}%
\emph{where on the rhs }$\frac{\partial M_{(g)\nu }^{\mu }(r^{\prime
},r(r^{\prime }))}{\partial r^{\prime \alpha }}$\emph{\ denotes the "total"
partial derivative with respect to} $r^{\prime \alpha },$ \emph{namely
defined such that the differential }$dM_{(g)\nu }^{\mu }(r^{\prime
},r(r^{\prime }))$\emph{\ is written explicitly as}%
\begin{equation}
dM_{(g)\nu }^{\mu }(r^{\prime },r)=dr^{\prime \alpha }\left[ \left. \frac{%
\partial M_{(g)\nu }^{\mu }(r^{\prime },r(r^{\prime }))}{\partial r^{\prime
\alpha }}\right\vert _{r(r^{\prime })}+\left. \frac{\partial M_{(g)\nu
}^{\mu }(\left( r^{\prime }\right) ,r(r^{\prime }))}{\partial r^{\prime
\alpha }}\right\vert _{r^{\prime }}\right] ,  \label{IDD-5}
\end{equation}%
\emph{where the partial derivatives on the rhs of the previous equation are
performed respectively at constant }$r(r^{\prime })$\emph{\ the first one
and at constant }$r^{\prime }$\emph{\ the other one.}

\emph{Proof -} In fact, thanks to Eq.(\ref{IDD-3}), the partial derivative
of the Jacobian $M_{(g)\nu }^{\mu }(r^{\prime },r)$ with respect to $%
r^{\prime \alpha }$ becomes%
\begin{equation}
\frac{\partial }{\partial r^{\prime \alpha }}M_{(g)\nu }^{\mu }(r^{\prime
},r)=\left. \frac{\partial M_{(g)\nu }^{\mu }(r^{\prime },r)}{\partial
r^{\prime \alpha }}\right\vert _{r}+\frac{\partial r^{\beta }}{\partial
r^{\prime \alpha }}\left. \frac{\partial M_{(g)\nu }^{\mu }(r^{\prime },r)}{%
\partial r^{\beta }}\right\vert _{r^{\prime }},
\end{equation}%
while its differential is just%
\begin{eqnarray}
dM_{(g)\nu }^{\mu }(r^{\prime },r) &=&dr^{\prime \alpha }\left. \frac{%
\partial M_{(g)\nu }^{\mu }(r^{\prime },r)}{\partial r^{\prime \alpha }}%
\right\vert _{r}+dr^{\beta }\left. \frac{\partial M_{\nu }^{\mu }(r^{\prime
},r)}{\partial r^{\beta }}\right\vert _{r^{\prime }}  \notag \\
&=&dr^{\prime \alpha }\left[ \left. \frac{\partial M_{(g)\nu }^{\mu
}(r^{\prime },r)}{\partial r^{\prime \alpha }}\right\vert _{r}+M_{(g)\alpha
}^{\beta }(r^{\prime },r)\left. \frac{\partial M_{(g)\nu }^{\mu }(r^{\prime
},r)}{\partial r^{\beta }}\right\vert _{r^{\prime }}\right] .
\end{eqnarray}%
Hence due to Eq.(\ref{IDD-3}) it follows that%
\begin{eqnarray}
dM_{(g)\nu }^{\mu }(r^{\prime },r) &=&dr^{\prime \alpha }\left. \frac{%
\partial M_{(g)\nu }^{\mu }(r^{\prime },r)}{\partial r^{\prime \alpha }}%
\right\vert _{r}+dr^{\beta }\left. \frac{\partial M_{(g)\nu }^{\mu
}(r^{\prime },r)}{\partial r^{\beta }}\right\vert _{r^{\prime }}  \notag \\
&=&dr^{\prime \alpha }\left[ \left. \frac{\partial M_{(g)\nu }^{\mu
}(r^{\prime },r(r^{\prime }))}{\partial r^{\prime \alpha }}\right\vert
_{r(r^{\prime })}+M_{(g)\alpha }^{\beta }(r^{\prime },r)\left(
M_{(g)}^{-1}\right) _{\alpha }^{\beta }\left. \frac{\partial M_{(g)\nu
}^{\mu }(r^{\prime },r(r^{\prime }))}{\partial r^{\prime \beta }}\right\vert
_{r^{\prime }}\right] ,
\end{eqnarray}%
which manifestly implies Eq.(\ref{IDD-5}). The rhs of the same equation
coincides then with the rhs of Eq.(\ref{IDD-4}) so that the thesis is
proved. \textbf{Q.E.D.}

Let us now inspect further mathematical implications which are implied in
the construction of the NLPTs.

\section{3 - NLPT-transformation properties of the Christoffel symbols}

In this section we intend to determine the transformations properties of the
Christoffel symbols with respect to the general-NLPT-group $\left\{
P_{g}\right\} $. For definiteness, let us denote as $\Gamma _{\alpha \beta
}^{\nu }$ and $\Gamma _{\alpha \beta }^{\prime \nu }$ the (initial and
transformed) Christoffel symbols when referred to the two GR-reference
frames $r^{\mu }$ ("initial frame") and $r^{\prime \mu }$ ("transformed
frame") respectively. The issue is the determination of the relationship
between $\Gamma _{\alpha \beta }^{\nu }$ and $\Gamma _{\alpha \beta
}^{\prime \nu }$ when the coordinate transformation relating the coordinates
$r^{\mu }$ and $r^{\prime \mu }$ is suitably prescribed.

As we intend to show here, the solution of such a problem is closely related
to the requirement, already included in the prescription of the group of
NLPT $\left\{ P_{g}\right\} $, that the (initial and transformed) metric
tensors $g^{ij}(r)$ and $g^{\prime ij}(r^{\prime })$ defined respectively
for the two Riemannian manifolds $\left\{ \mathbf{Q}^{4},g\right\} $ and $%
\left\{ \mathbf{Q}^{\prime 4},g^{\prime }\right\} $ are extremal, namely
they satisfy identically the \emph{extremal conditions}%
\begin{eqnarray}
\nabla _{k}g^{ij}(r) &=&0,  \label{CONTRAINT-1} \\
\nabla _{k}^{\prime }g^{\prime ij}(r^{\prime }) &=&0.  \label{CONTRAINT-2}
\end{eqnarray}%
Here $\nabla _{j}$ and $\nabla _{j}^{\prime }$ denote as usual the covariant
derivatives, defined as%
\begin{eqnarray}
\nabla _{k}g^{ij}(r) &=&\frac{\partial g^{ij}(r)}{\partial r^{k}}+\Gamma
_{kl}^{i}g^{lj}(r)+\Gamma _{kl}^{j}g^{il}(r),  \label{def-1} \\
\nabla _{k}^{\prime }g^{\prime ij}(r^{\prime }) &=&\frac{\partial g^{\prime
ij}(r^{\prime })}{\partial r^{\prime k}}+\Gamma _{kl}^{\prime i}g^{\prime
lj}(r^{\prime })+\Gamma _{kl}^{\prime j}g^{\prime il}(r^{\prime }),
\label{def-2}
\end{eqnarray}%
where $\Gamma _{kl}^{i}$ and $\Gamma _{kl}^{\prime i}$ denote the initial
and transformed Christoffel symbols defined on $\left\{ \mathbf{Q}%
^{4},g\right\} $ and $\left\{ \mathbf{Q}^{\prime 4},g^{\prime }\right\} $
respectively. We notice that Eqs.(\ref{CONTRAINT-1}) and (\ref{CONTRAINT-2})
are manifestly equivalent to the ODE's%
\begin{eqnarray}
\frac{D}{Ds}g^{ij}(r(s)) &=&0,  \label{CONSTRAINT-1A} \\
\frac{D^{\prime }}{Ds}g^{\prime ij}(r^{\prime }(s)) &=&0,
\label{CONSTRAINT-2A}
\end{eqnarray}%
once $\frac{D}{Ds}$ and $\frac{D^{\prime }}{Ds}$ are identified with the
\emph{covariant derivatives} of $g^{ij}(r(s)$ and $g^{\prime ij}(r^{\prime
}(s))$. These are defined respectively in $\mathbf{Q}^{4}$ and $\mathbf{Q}%
^{\prime 4}$ in terms of the differential operators acting on the covariants
components $g^{ij}(r(s))$ and $g^{\prime ij}(r^{\prime }(s))$ as%
\begin{eqnarray}
\frac{D}{Ds}g^{\prime ij}(r^{\prime }(s) &=&u^{k}\nabla _{k}g^{ij}(r),
\label{DIFF-OP-1} \\
\frac{D^{\prime }}{Ds}g^{\prime ij}(r^{\prime }) &=&u^{\prime k}\nabla
_{k}^{\prime }g^{\prime ij}(r^{\prime }),  \label{DIFF-OP-2}
\end{eqnarray}%
with $u^{k}$ and $u^{\prime k}=\left( M^{-1}\right) _{j}^{k}u^{\prime }$
denoting the $4-$velocities in the corresponding tangent spaces. Then the
following proposition holds.

\bigskip

\textbf{THM.1 - NLPT-transformation laws for the Christoffel symbols}

\emph{Within the group }$\left\{ P_{g}\right\} $\emph{\ the following two
propositions hold:}

L$_{1})$ \emph{The extremal conditions (\ref{CONTRAINT-1}) and (\ref%
{CONTRAINT-2}) holding for the metric tensors }$g^{ij}(r)$\emph{\ and }$%
g^{\prime ij}(r^{\prime })$\emph{\ are equivalent to require that the
initial and transformed Christoffel symbols }$\Gamma _{\nu \gamma }^{k}$%
\emph{\ and }$\Gamma _{\nu \gamma }^{\prime k}$\emph{\ defined respectively
on }$\left\{ \mathbf{Q}^{4},g\right\} $\emph{\ and }$\left\{ \mathbf{Q}%
^{\prime 4},g^{\prime }\right\} $\emph{\ satisfy the constraint differential
equation}%
\begin{equation}
dr^{\prime s}\frac{\partial }{\partial r^{\prime s}}M_{(g)\nu }^{\mu
}(r^{\prime },r)+\Gamma _{\alpha \beta }^{\mu }M_{(g)\nu }^{\alpha
}(r^{\prime },r)M_{(g)s}^{\beta }(r^{\prime },r)dr^{\prime s}=M_{(g)\gamma
}^{\mu }(r^{\prime },r)\Gamma _{\nu s}^{\prime \gamma }dr^{\prime s}.
\label{Lemma-3-1}
\end{equation}

L$_{2})$ \emph{The previous equation in turn is equivalent to the equation}%
\begin{equation}
\Gamma _{\nu \gamma }^{\prime k}=\left( M_{(g)}^{-1}\right) _{\mu
}^{k}(r,r^{\prime })\frac{\partial M_{(g)\nu }^{\mu }(r^{\prime },r)}{%
\partial r^{\prime \gamma }}+\left( M_{(g)}^{-1}\right) _{\mu
}^{k}(r,r^{\prime })\Gamma _{\alpha \beta }^{\mu }M_{(g)\nu }^{\alpha
}(r^{\prime },r)M_{(g)\gamma }^{\beta }(r^{\prime },r),  \label{Lemma-3-2}
\end{equation}%
\emph{which determines the transformation laws for the transformed
Christoffel symbol} $\Gamma _{\nu \gamma }^{\prime k}.$

\emph{Proof -} We first prove proposition L$_{2})$. Notice for this purpose
that Eq.(\ref{Lemma-3-1}), due to the arbitrariness of the differential
displacement $dr^{\prime s}$, implies also that%
\begin{equation}
\frac{\partial }{\partial r^{\prime s}}M_{(g)\nu }^{\mu }(r^{\prime
},r)+\Gamma _{\alpha \beta }^{\mu }M_{\nu }^{\alpha }(r^{\prime
},r)M_{(g)s}^{\beta }(r^{\prime },r)=M_{(g)\gamma }^{\mu }(r^{\prime
},r)\Gamma _{\nu s}^{\prime \gamma },  \label{Lemma-3-3}
\end{equation}%
which, after multiplying it term by term by $\left( M_{(g)}^{-1}\right)
_{\mu }^{k}(r,r^{\prime })$, exchanging the indexes $s\leftrightarrow \gamma
$ and recalling that $M_{(g)s}^{\mu }(r,r^{\prime })\left(
M_{(g)}^{-1}\right) _{\mu }^{k}(r,r^{\prime })=\delta _{s}^{k}$, reduces to
Eq.(\ref{Lemma-3-2}).

Next we address the proof of proposition L$_{1})$, i.e., that Eq.(\ref%
{CONTRAINT-1}) is equivalent to Eq.(\ref{Lemma-3-2}). To start with, let us
consider the definition of the covariant derivative recalled above [see Eq.(%
\ref{def-1})]. Then, Eq.(\ref{CONTRAINT-1}) delivers necessarily:%
\begin{eqnarray}
\nabla _{k}g^{ij}(r) &=&\frac{\partial g^{ij}(r)}{\partial r^{k}}+\Gamma
_{kl}^{i}g^{lj}(r)+\Gamma _{kl}^{j}g^{il}(r)=  \notag \\
&=&\frac{\partial }{\partial r^{k}}\left[ M_{(g)\alpha }^{i}\left( r^{\prime
},r\right) M_{(g)\beta }^{j}\left( r^{\prime },r\right) g^{\prime \alpha
\beta }\left( r^{\prime }\right) \right] +  \notag \\
&&+\Gamma _{kl}^{i}M_{(g)\alpha }^{l}\left( r^{\prime },r\right) M_{(g)\beta
}^{j}\left( r^{\prime },r\right) g^{\prime \alpha \beta }\left( r^{\prime
}\right) +  \notag \\
&&\left. +\Gamma _{kl}^{j}M_{(g)\alpha }^{i}\left( r^{\prime },r\right)
M_{(g)\beta }^{l}\left( r^{\prime },r\right) g^{\prime \alpha \beta }\left(
r^{\prime }\right) =0.\right.  \label{alfa}
\end{eqnarray}%
Invoking now the identity%
\begin{equation}
\frac{\partial }{\partial r^{k}}g^{\prime \alpha \beta }\left( r^{\prime
}\right) =\left( M_{(g)}^{-1}\right) _{k}^{s}(r,r^{\prime })\frac{\partial }{%
\partial r^{\prime s}}g^{\prime \alpha \beta }\left( r^{\prime }\right) ,
\end{equation}%
thanks to the chain rule, it follows that the first term on the rhs of Eq.(%
\ref{alfa}) becomes%
\begin{eqnarray}
\frac{\partial }{\partial r^{k}}\left[ M_{(g)\alpha }^{i}\left( r^{\prime
},r\right) M_{(g)\beta }^{j}\left( r^{\prime },r\right) g^{\prime \alpha
\beta }\left( r^{\prime }\right) \right] &=&g^{\prime \alpha \beta }\left(
r^{\prime }\right) \frac{\partial }{\partial r^{k}}\left[ M_{(g)\alpha
}^{i}\left( r^{\prime },r\right) M_{(g)\beta }^{j}\left( r^{\prime
},r\right) \right]  \notag \\
&&+M_{(g)\alpha }^{i}\left( r^{\prime },r\right) M_{(g)\beta }^{j}\left(
r^{\prime },r\right) \left( M_{(g)}^{-1}\right) _{k}^{s}(r,r^{\prime })\frac{%
\partial }{\partial r^{\prime s}}g^{\prime \alpha \beta }\left( r^{\prime
}\right) .
\end{eqnarray}%
Therefore, noting that thanks to Eqs.(\ref{CONTRAINT-2}) it must be%
\begin{equation}
\frac{\partial }{\partial r^{\prime s}}g^{\prime \alpha \beta }\left(
r^{\prime }\right) =\nabla _{s}^{\prime }g^{\prime \alpha \beta }(r^{\prime
})-\Gamma _{sl}^{\prime \alpha }g^{\prime l\beta }(r^{\prime })+\Gamma
_{sl}^{\prime \beta }g^{\prime \alpha l}(r^{\prime }),
\end{equation}%
it follows that Eq.(\ref{alfa}) requires necessarily the validity of the
following constraint equation, obtained also upon exchanging summations
indexes, namely%
\begin{eqnarray}
&&g^{\prime \alpha \beta }\left( r^{\prime }\right) M_{(g)\alpha }^{i}\left(
r^{\prime },r\right) \frac{\partial }{\partial r^{k}}M_{(g)\beta }^{j}\left(
r^{\prime },r\right) +g^{\prime \alpha \beta }\left( r^{\prime }\right)
M_{(g)\beta }^{j}\left( r^{\prime },r\right) \frac{\partial }{\partial r^{k}}%
M_{(g)\alpha }^{i}\left( r^{\prime },r\right) -  \notag \\
&&-M_{(g)q}^{i}\left( r^{\prime },r\right) M_{(g)\beta }^{j}\left( r^{\prime
},r\right) \left( M_{(g)}^{-1}\right) _{k}^{s}(r,r^{\prime })\Gamma
_{s\alpha }^{\prime q}g^{\prime \alpha \beta }(r^{\prime })+  \notag \\
&&-M_{(g)q}^{i}\left( r^{\prime },r\right) M_{(g)p}^{j}\left( r^{\prime
},r\right) \left( M_{(g)}^{-1}\right) _{k}^{s}(r,r^{\prime })\Gamma _{s\beta
}^{\prime p}g^{\prime \alpha \beta }(r^{\prime })+  \notag \\
&&+\Gamma _{kl}^{i}M_{(g)\alpha }^{l}\left( r^{\prime },r\right) M_{(g)\beta
}^{j}\left( r^{\prime },r\right) g^{\prime \alpha \beta }\left( r^{\prime
}\right) +  \notag \\
&&\left. +\Gamma _{kl}^{j}M_{(g)\alpha }^{i}\left( r^{\prime },r\right)
M_{(g)\beta }^{l}\left( r^{\prime },r\right) g^{\prime \alpha \beta }\left(
r^{\prime }\right) =0.\right.  \label{last}
\end{eqnarray}%
Considering now $g^{\prime \alpha \beta }\left( r^{\prime }\right) $ as
independent of the Jacobian matrix of the transformation, then thanks to the
symmetry of the indexes $\alpha $ and $\beta $\ the previous equation
delivers%
\begin{equation}
M_{(g)\beta }^{j}\left( r^{\prime },r\right) \left[ \frac{\partial }{%
\partial r^{k}}M_{(g)\alpha }^{i}\left( r^{\prime },r\right)
-M_{q}^{i}\left( r^{\prime },r\right) \left( M^{-1}\right)
_{k}^{s}(r,r^{\prime })\Gamma _{s\alpha }^{\prime q}\right] +\Gamma
_{kl}^{i}M_{\alpha }^{l}\left( r^{\prime },r\right) M_{\beta }^{j}\left(
r^{\prime },r\right) =0.
\end{equation}%
Namely, multiplying term by term by $\left( M^{-1}\right)
_{j}^{s}(r,r^{\prime })$%
\begin{eqnarray}
&&\delta _{\beta }^{s}\left[ \frac{\partial }{\partial r^{k}}M_{\alpha
}^{i}\left( r^{\prime },r\right) -M_{(g)q}^{i}\left( r^{\prime },r\right)
\left( M_{(g)}^{-1}\right) _{k}^{s}(r,r^{\prime })\Gamma _{s\alpha }^{\prime
q}\right]  \notag \\
&&\left. +\Gamma _{kl}^{i}M(g)_{\alpha }^{l}\left( r^{\prime },r\right)
M_{(g)\beta }^{j}\left( r^{\prime },r\right) \left( M_{(g)}^{-1}\right)
_{j}^{s}(r,r^{\prime })=0,\right.
\end{eqnarray}%
which yields%
\begin{equation}
\delta _{\beta }^{s}\left[ \frac{\partial }{\partial r^{k}}M_{(g)\alpha
}^{i}\left( r^{\prime },r\right) -M_{(g)q}^{i}\left( r^{\prime },r\right)
\left( M_{(g)}^{-1}\right) _{k}^{r}(r,r^{\prime })\Gamma _{r\alpha }^{\prime
q}+\Gamma _{kl}^{i}M_{(g)\alpha }^{l}\left( r^{\prime },r\right) \right] =0.
\end{equation}%
Therefore, one has that%
\begin{equation}
\frac{\partial }{\partial r^{k}}M_{(g)\alpha }^{i}\left( r^{\prime
},r\right) -M_{(g)q}^{i}\left( r^{\prime },r\right) \left(
M_{(g)}^{-1}\right) _{k}^{r}(r,r^{\prime })\Gamma _{r\alpha }^{\prime
q}+\Gamma _{kl}^{i}M_{(g)\alpha }^{l}\left( r^{\prime },r\right) =0,
\end{equation}%
implying also%
\begin{equation}
\left( M_{(g)}^{-1}\right) _{k}^{r}(r,r^{\prime })\frac{\partial }{\partial
r^{\prime r}}M_{(g)\alpha }^{i}\left( r^{\prime },r\right)
-M_{(g)q}^{i}\left( r^{\prime },r\right) \left( M_{(g)}^{-1}\right)
_{k}^{r}(r,r^{\prime })\Gamma _{r\alpha }^{\prime q}+\Gamma
_{kl}^{i}M_{(g)\alpha }^{l}\left( r^{\prime },r\right) =0.
\end{equation}%
Hence it follows%
\begin{eqnarray}
&&M_{(g)x}^{k}\left( r^{\prime },r\right) \left( M_{(g)}^{-1}\right)
_{k}^{r}(r,r^{\prime })\frac{\partial }{\partial r^{\prime r}}M_{(g)\alpha
}^{i}\left( r^{\prime },r\right) -M_{(g)q}^{i}\left( r^{\prime },r\right)
M_{(g)x}^{k}\left( r^{\prime },r\right) \left( M_{(g)}^{-1}\right)
_{k}^{r}(r,r^{\prime })\Gamma _{r\alpha }^{\prime q}+  \notag \\
&&\left. +M_{(g)x}^{k}\left( r^{\prime },r\right) \Gamma
_{kl}^{i}M_{(g)\alpha }^{l}\left( r^{\prime },r\right) =0,\right.
\end{eqnarray}%
so that%
\begin{equation}
\delta _{x}^{r}\frac{\partial }{\partial r^{\prime r}}M_{(g)\alpha
}^{i}\left( r^{\prime },r\right) -M_{(g)q}^{i}\left( r^{\prime },r\right)
\delta _{x}^{r}\Gamma _{r\alpha }^{\prime q}+M_{(g)x}^{k}\left( r^{\prime
},r\right) \Gamma _{kl}^{i}M_{(g)\alpha }^{l}\left( r^{\prime },r\right) =0.
\end{equation}%
Finally, replacing the index $x$ with $k$ one gets%
\begin{equation}
\frac{\partial }{\partial r^{\prime k}}M_{(g)\alpha }^{i}\left( r^{\prime
},r\right) -M_{(g)q}^{i}\left( r^{\prime },r\right) \Gamma _{k\alpha
}^{\prime q}+M_{(g)k}^{p}\left( r^{\prime },r\right) M_{(g)\alpha
}^{q}\left( r^{\prime },r\right) \Gamma _{pq}^{i}=0.
\end{equation}

Straightforward algebra shows that this equation coincides with Eq.(\ref%
{Lemma-3-3}) and hence Eq.(\ref{Lemma-3-1}) too. Finally, one can show that
in a similar way Eq.(\ref{Lemma-3-2}) implies Eq.(\ref{CONTRAINT-1}) too. In
view of the equivalence between Eqs. (\ref{Lemma-3-2}), (\ref{Lemma-3-1})
and (\ref{Lemma-3-2}) the thesis is reached. \textbf{Q.E.D.}

\bigskip

The implication of THM.1 is therefore that the requirements that both the
initial and transformed metric tensors are extremal, i.e., in the sense that
the corresponding covariant derivatives vanish identically in both cases
(see Eqs.(\ref{CONSTRAINT-1A}) and (\ref{CONSTRAINT-2A})), is necessarily
equivalent to impose between the initial and transformed Christoffel symbols
- i.e., $\Gamma _{\nu \gamma }^{k}$\emph{\ and }$\Gamma _{\nu \gamma
}^{\prime k}$\emph{\ which are defined respectively on }$\left\{ \mathbf{Q}%
^{4},g\right\} $\emph{\ and }$\left\{ \mathbf{Q}^{\prime 4},g^{\prime
}\right\} $ -\emph{\ }the transformation law (\ref{Lemma-3-2}). The
conclusion, as shown in the next section, is important to establish the
tensor transformation laws which hold for the $4-$acceleration for arbitrary
NLPTs belonging to the group $\left\{ P_{g}\right\} $.

As a final comment, we remark that if the space-time $\left\{ \mathbf{Q}%
^{\prime 4},g^{\prime }\right\} $ is identified with the flat Lorentzian
Minkowski space-time $\left\{ \mathbf{M}^{4},\eta \right\} $\emph{\ }then
the following proposition holds.

\bigskip

\textbf{COROLLARY TO THM.1 - Case of Minkowski space-time}

\emph{Within the group }$\left\{ P_{g}\right\} $\emph{\ if the space-time} $%
\left\{ \mathbf{Q}^{\prime 4},g^{\prime }\right\} $ \emph{coincides with the
flat Lorentzian Minkowski space-time }$\left\{ \mathbf{M}^{4},\eta \right\} $%
\emph{\ expressed in orthogonal Cartesian coordinates, then Eq.(\ref%
{Lemma-3-2}) reduces to}%
\begin{equation}
0=\frac{\partial M_{(g)\nu }^{\mu }(r^{\prime },r)}{\partial r^{\prime
\gamma }}+\Gamma _{\alpha \beta }^{\mu }M_{(g)\nu }^{\alpha }(r^{\prime
},r)M_{(g)\gamma }^{\beta }(r^{\prime },r),  \label{Corollary to THM.2}
\end{equation}%
\emph{which provides a representation for the Christoffel symbol} $\Gamma
_{\alpha \beta }^{\mu }.$

\emph{Proof} - Assume in fact that the space-time $\left\{ \mathbf{Q}%
^{\prime 4},g^{\prime }\right\} $ coincides with the flat Minkowski
space-time $\left\{ \mathbf{M},\eta \right\} .$ Then by construction in Eq.(%
\ref{Lemma-3-2}) the transformed Christoffel symbols $\Gamma _{\alpha \beta
}^{\prime \mu }$ in such a space-time necessarily vanish identically. Then
the same equation reduces to Eq.(\ref{Corollary to THM.2}). \textbf{Q.E.D.}

\section{4 - NLPT-transformation properties of the $4-$acceleration}

THM.1 and in particular Eq.(\ref{Lemma-3-2}) can be used to determine also
the relationships holding between the $4-$accelerations defined in the two
Riemannian manifolds $\left\{ \mathbf{Q}^{4},g\right\} $ and $\left\{
\mathbf{Q}^{\prime 4},g^{\prime }\right\} $ respectively. In fact, let us
identify the $4-$accelerations with the covariant derivatives of $u^{\nu }$
and $u^{\prime \nu }$ defined in $\mathbf{Q}^{4}$ and $\mathbf{Q}^{\prime 4}$
as%
\begin{eqnarray}
a^{\mu } &\equiv &\frac{D}{Ds}u^{\mu },  \label{4-ac-1} \\
a^{\prime \mu } &\equiv &\frac{D^{\prime }}{Ds}u^{\prime \mu },
\label{4-acc-2}
\end{eqnarray}%
and where $\frac{D}{Ds}$ and $\frac{D^{\prime }}{Ds}$ are identified with
the ordinary differential operators (\ref{DIFF-OP-1}) and (\ref{DIFF-OP-2}).
This means that in the two space-times they must be identified respectively
as%
\begin{equation}
\frac{D}{Ds}u^{\nu }=\frac{d}{ds}u^{\nu }+u^{\alpha }u^{\beta }\Gamma
_{\alpha \beta }^{\nu },  \label{T3-1}
\end{equation}%
\begin{equation}
\frac{D^{\prime }}{Ds}u^{\prime \nu }=\frac{d}{ds}u^{\prime \nu }+u^{\prime
\alpha }u^{\prime \beta }\Gamma _{\alpha \beta }^{\prime \nu },  \label{T3-2}
\end{equation}%
where $\Gamma _{\alpha \beta }^{\nu }$ and $\Gamma _{\alpha \beta }^{\prime
\nu }$ denote the corresponding standard connections defined in the same
space-times. Let us consider for definiteness Eq.(\ref{T3-1}), the other one
being uniquely dependent from it (as will be obvious from the subsequent
considerations). Invoking the tensor transformation laws for the $4-$%
velocity (\ref{4-velocity trasnformations}) then Eq.(\ref{T3-1}) implies that%
\begin{equation}
\frac{d}{ds}u^{\mu }=M_{(g)\nu }^{\prime \prime \mu }\frac{D}{Ds}u^{\prime
\nu }-M_{(g)\nu }^{\prime \mu }u^{\prime \alpha }u^{\prime \beta }\Gamma
_{\alpha \beta }^{\prime \nu }+u^{\prime \nu }\frac{d}{ds}M_{(g)\nu
}^{\prime \mu }.  \label{D1}
\end{equation}%
Then it is immediate to prove that, thanks to the validity of THM.1, the $4-$%
accelerations $\frac{D}{Ds}u^{\nu }$ and $\frac{D^{\prime }}{Ds}u^{\prime
\nu }$ are linearly related. In particular the following result holds.

\bigskip

\textbf{THM.2 - NLPT-transformation law for the 4-acceleration}

\emph{If }$M_{\nu }^{\prime \mu }$\emph{\ is the Jacobian of an arbitrary
NLPT, defined according to Eq.(\ref{poi5}), and }$\frac{D}{Ds}u^{\nu }$\emph{%
\ and }$\frac{D^{\prime }}{Ds}u^{\prime \nu }$\emph{\ are the }$\emph{4-}$%
\emph{accelerations defined according to Eqs.(\ref{T3-1}) and (\ref{T3-2}),
then with respect to an arbitrary NLPT of the group }$\left\{ P_{g}\right\} $%
\emph{\ it follows that they are related by means of the tensor
transformation laws}%
\begin{equation}
\frac{D}{Ds}u^{\mu }=M_{(g)\nu }^{\mu }(r^{\prime },r)\frac{D^{\prime }}{Ds}%
u^{\prime \nu },  \label{T3-3}
\end{equation}%
\begin{equation}
\frac{D^{\prime }}{Ds}u^{\prime \mu }=\left( M_{(g)}^{-1}\right) _{\nu
}^{\mu }(r,r^{\prime })\frac{D}{Ds}u^{\nu }.  \label{T3-4}
\end{equation}

\emph{The result is analogous to that holding for arbitrary LPTs belonging
to the group }$\left\{ P_{g}\right\} .$

\emph{Proof -} First it is obvious that Eqs.(\ref{T3-3}) and (\ref{T3-4})
mutually imply each other so that it is sufficient to prove that one of the
two actually holds. Consider then the proof of Eq.(\ref{T3-3}). First, let
us invoke the transformation law for the $4-$velocity (\ref{4-velocity
trasnformations}) and invoke Eq.(\ref{T3-1}) to give%
\begin{equation}
\frac{D}{Ds}u^{\mu }=\frac{D}{Ds}\left[ M_{(g)\nu }^{\mu }(r^{\prime
},r)u^{\prime \nu }\right] =\frac{d}{ds}\left[ M_{(g)\nu }^{\mu }(r^{\prime
},r)u^{\prime \nu }\right] +u^{\prime h}u^{\prime k}M_{(g)h}^{\alpha
}(r^{\prime },r)M_{(g)k}^{\beta }(r^{\prime },r)\Gamma _{\alpha \beta }^{\nu
},
\end{equation}%
where the chain rule delivers%
\begin{equation*}
\frac{d}{ds}\left[ M_{(g)\nu }^{\mu }(r^{\prime },r)u^{\prime \nu }\right]
=M_{(g)\nu }^{\mu }(r^{\prime },r)\frac{d}{ds}u^{\prime \nu }+u^{\prime \nu }%
\frac{d}{ds}\left[ M_{(g)\nu }^{\mu }(r^{\prime },r)\right] .
\end{equation*}%
Its is immediate to show that Lemma and Eq.(\ref{Lemma-3-2}) of THM.1 then
imply identically the identity%
\begin{equation}
dM_{(g)\nu }^{\mu }(r^{\prime },r)+\Gamma _{\alpha \beta }^{\mu }M_{(g)\nu
}^{\alpha }(r^{\prime },r)M_{(g)\gamma }^{\beta }(r^{\prime },r)dr^{\prime
\gamma }=M_{(g)\gamma }^{\mu }(r^{\prime },r)\Gamma _{\nu \beta }^{\prime
\gamma }dr^{\prime \beta }.
\end{equation}

Hence the thesis is proved. Incidentally, thanks to Lemma 1, it is obvious
that the same conclusion holds in the case of arbitrary LPTs belonging to
group $\left\{ P\right\} .$ \textbf{Q.E.D.}

\section{5 - Application \#1: NLPT-acceleration effects in Schwarzschild,
Reissner-Nordstr\"{o}m\emph{\ }and Schwarzschild-analog space-times}

The first application to be considered concerns the construction of a NLPT
mapping two connected and time-oriented space-times $(\mathbf{Q}^{4},g)$\
and $(\mathbf{Q}^{\prime 4},g^{\prime })$ both having diagonal form with
respect to suitable sets of coordinates. More precisely we shall require
that:

\begin{itemize}
\item When $(\mathbf{Q}^{4},g)$\ and $(\mathbf{Q}^{\prime 4},g^{\prime })$
are referred to the same coordinate systems both are realized by diagonal
metric tensors
\begin{equation}
\left\{
\begin{array}{c}
g_{\mu \nu }(r)\equiv diag\left(
S_{0}(r),-S_{1}(r),-S_{2}(r),-S_{3}(r)\right) \\
g_{\mu \nu }^{\prime }(r^{\prime })\equiv diag\left( S_{0}^{\prime
}(r^{\prime }),-S_{1}^{\prime }(r^{\prime }),-S_{2}^{\prime }(r^{\prime
}),-S_{3}^{\prime }(r^{\prime })\right)%
\end{array}%
\right.  \label{diagonal metric tensors}
\end{equation}%
\ respectively.\ The accessible subsets are as follows: a) for $(\mathbf{Q}%
^{\prime 4},g^{\prime })$ this is that in which for all $\mu =0,3,$ $S_{\mu
}^{\prime }(r^{\prime })>0;$ b) for $(\mathbf{Q}^{4},g)$ is either the set
in which for all $\mu =0,3,$ $S_{\mu }(r)>0$ or the other one in which $%
S_{0}(r)<0,S_{1}(r)<0,S_{2}(r)>0$ and $S_{3}(r)>0.$

\item $(\mathbf{Q}^{4},g)$\ and $(\mathbf{Q}^{\prime 4},g^{\prime })$ are
intrinsically \emph{different}, i.e., that the corresponding Riemann
curvature tensors $R_{\mu \nu }(r)$\ and $R_{\mu \nu }^{\prime }(r^{\prime
}) $\ cannot be globally mapped in each other by means of any LPT. This
means that a mapping between the accessible subsets of the said space-times
can only possibly be established by means of a suitable NLPT.

\item We shall consider for definiteness only the case in which both $(%
\mathbf{Q}^{\prime 4},g^{\prime })$ and $(\mathbf{Q}^{4},g)$ have the same
Lorentzian signature $(+,-,-,-).$
\end{itemize}

In validity of Eqs.(\ref{diagonal metric tensors}) the tensor transformation
equation for the metric tensor (see Eq. (24) in Part 2) take the general
form:%
\begin{equation}
\left\{
\begin{tabular}{l}
$S_{\mu }(r)=\left( M_{(g)}^{-1}\right) _{\mu }^{\alpha }(r,r^{\prime
})\left( M_{(g)}^{-1}\right) _{(\mu )}^{\alpha }(r,r^{\prime })S_{\alpha
}^{\prime }(r^{\prime }),$ \\
$S_{\alpha }^{\prime }(r^{\prime })=M_{(g)\mu }^{\alpha }(r^{\prime
},r)M_{(g)(\mu )}^{\alpha }(r^{\prime },r)S_{\alpha }(r),$%
\end{tabular}%
\right.  \label{diagonal-case-SOLUTION}
\end{equation}%
where manifestly $M_{(g)\mu }^{\alpha }(r^{\prime },r)\equiv M_{\mu
}^{\alpha }(r^{\prime },r)$ and $\left( M_{(g)}^{-1}\right) _{\mu }^{\alpha
}(r,r^{\prime })=\left( M^{-1}\right) _{\mu }^{\alpha }(r,r^{\prime })$ as
corresponds to the case of a special NLTP. For such a type of space-times in
the following we intend to display a number of explicit particular solutions
of Eqs.(\ref{diagonal-case-SOLUTION}) for the Jacobian $M_{\mu }^{\alpha }$
and its inverse $\left( M^{-1}\right) _{\mu }^{\alpha }$, and to construct
also the corresponding NLPT-phase-space transformations.

In the case in which $(\mathbf{Q}^{4},g)$\ and $(\mathbf{Q}^{\prime
4},g^{\prime })$ have the same signatures a particular solution of Eqs.(\ref%
{diagonal-case-SOLUTION}) in the accessible subsets of $(\mathbf{Q}^{4},g)$\
and $(\mathbf{Q}^{\prime 4},g^{\prime })$ is provided, as shown in Part 2,
by a diagonal Jacobian matrix, i.e., of the form%
\begin{equation}
M_{\mu }^{\alpha }(r^{\prime },r)=M_{\mu }^{\mu }(r^{\prime },r)\delta _{\mu
}^{\alpha }\equiv \left[ \delta _{\mu }^{\alpha }+A_{\mu }^{\mu }(r^{\prime
},r)\right] \delta _{\mu }^{\alpha },  \label{DIAGONAL MATRIX}
\end{equation}%
which corresponds to a \emph{diagonal NLPT} of the form%
\begin{equation}
\left\{
\begin{array}{c}
dr^{i}=M_{\left( i\right) }^{i}(r^{\prime },r)dr^{\prime \left( i\right) }
\\
\text{(}i=0,1,2,3\text{),}%
\end{array}%
\right.  \label{diagonal NLPT}
\end{equation}%
Indeed, from Eqs.(\ref{diagonal-case-SOLUTION}) one finds%
\begin{equation}
M_{(g)(\mu )}^{\mu }(r^{\prime },r)=\frac{1}{\left( M_{(g)}^{-1}\right)
_{(\mu )}^{\mu }}=\sqrt{\frac{S_{\mu }^{\prime }(r)}{S_{(\mu )}(r^{\prime })}%
},
\end{equation}%
where $\frac{S_{\mu }(r)}{S_{(\mu )}^{\prime }(r^{\prime })}>0$ in the
accessible subsets. In terms of Eqs.(\ref{T3-3}) and (\ref{T3-4}) one
obtains the acceleration transformation laws%
\begin{equation}
\left\{
\begin{array}{c}
\frac{D}{Ds}u^{\mu }=M_{I(g)(\mu )}^{\mu }(r^{\prime },r)\frac{D^{\prime }}{%
Ds}u^{\prime \mu } \\
\frac{D^{\prime }}{Ds}u^{\prime \mu }=\frac{1}{\left( M_{(g)}^{-1}\right)
_{(\mu )}^{\mu }}\frac{D}{Ds}u^{\nu }%
\end{array}%
\right. .  \label{acceleration-NLPT}
\end{equation}

Let us now consider a possible physical realizations for the space-times $(%
\mathbf{Q}^{4},g)$ and $(\mathbf{Q}^{\prime 4},g^{\prime })$ and the
corresponding metric tensors $g_{\mu \nu }(r)$\ and $g_{\mu \nu }^{\prime
}(r^{\prime })$ respectively. Here we consider the examples pointed out in
Part 2, namely concerning (A) Schwarzschild and (B) Reissner-Nordstr\"{o}m
space-times to be mapped onto either the (C) Minkowski or (D)
Schwartzchild-analog space-times. In pseudo-spherical coordinates *** $%
\left( r,r^{2},r^{3}\right) $ (see Part 1) the following generic
representation is assumed to hold for all of them of the form $g_{\mu \nu
}(r)\equiv diag\left( \left( S_{0}(r),-S_{1}(r),-S_{2}(r),-S_{3}(r)\right)
\right) $,%
\begin{equation}
\left\{
\begin{array}{c}
S_{0}(r)=a(r), \\
S_{1}(r)=b(r), \\
S_{2}(r)=g(r), \\
S_{3}(r)=g(r).%
\end{array}%
\right.  \label{DIAGONAL CASE}
\end{equation}%
In particular, the Schwarzschild, Reissner-Nordstr\"{o}m and
Schwartzchild-analog cases are obtained letting:%
\begin{equation}
\left\{
\begin{array}{c}
a(r)=f(r) \\
b(r)=\frac{1}{f(r)} \\
g(r)=1%
\end{array}%
\right. ,  \label{SCHWARTZ-like-0}
\end{equation}%
where respectively%
\begin{equation}
\left\{
\begin{array}{c}
\begin{array}{ccc}
f(r)=\left( 1-\frac{r_{s}}{r}\right) &  & \text{case A}%
\end{array}
\\
\begin{array}{ccc}
f(r)=\left( 1-\frac{r_{s}}{r}+\frac{r_{Q}^{2}}{r^{2}}\right) &  & \text{case
B}%
\end{array}
\\
\begin{array}{ccc}
f(r)=\prod\limits_{i=1,n}(1-\frac{r_{i}}{r}) &  & \text{case C}%
\end{array}%
\end{array}%
\right. .  \label{SCHWARTZ-like-1}
\end{equation}%
Here, $r_{s},$ $r_{Q}$ and $r_{i}$ (for $i=1,n$) are suitably-prescribed
characteristic scale lengths, in particular
\begin{eqnarray}
r_{s} &=&\frac{2GM}{c^{2}},  \label{radius-1} \\
r_{Q} &=&\sqrt{\frac{Q^{2}G}{4\pi \varepsilon _{0}c^{4}}},  \label{radius-2}
\end{eqnarray}%
\ are respectively the Schwarzschild and Reissner-Nordstr\"{o}m\emph{\ }%
radii,\ with $Q$\ being the electric charge and $1/4\pi \varepsilon _{0}$\
the Coulomb coupling constant. In all cases A, B and C we shall require that
the function $f(r)$ defined according to Eq.(\ref{SCHWARTZ-like-1}) is
strictly positive, i.e., $r>r_{s}$ and $r>r_{n},$ with $r_{n}$ denoting the
largest root of the equation $f(r)=\left( 1-\frac{r_{s}}{r}+\frac{r_{Q}^{2}}{%
r^{2}}\right) =0$ or $\prod\limits_{i=1,n}(1-\frac{r_{i}}{r})=0.$
In all cases, the transformed space-time $(\mathbf{Q}^{\prime
4},g^{\prime })$\ when expressed in the same pseudo-spherical
coordinates is identified either with the Minkowski space-time or
with the Schwartzchild-analog space-time,
so that respectively either for all $\mu =0,3:$%
\begin{equation}
S_{\mu }^{\prime }(r^{\prime })=1,
\end{equation}%
or where (\ref{DIAGONAL CASE}), (\ref{SCHWARTZ-like-0}) and case C of Eq. (%
\ref{SCHWARTZ-like-1}) applies. In the subsets where $f(r)>0$ and for case C
$f^{\prime }(r^{\prime })>0$ the transformation matrix $M_{\mu }^{\nu
}(r^{\prime },r)$ becomes:%
\begin{equation}
\left\{
\begin{array}{c}
M_{0}^{0}(r^{\prime },r)=\sqrt{\frac{a^{\prime }(r^{\prime })}{a(r)}} \\
M_{1}^{1}(r^{\prime },r)=\sqrt{\frac{b^{\prime }(r^{\prime })}{b(r)}} \\
M_{2}^{2}(r^{\prime },r)=M_{3}^{3}(r^{\prime },r)=1%
\end{array}%
\right. ,  \label{acc-1}
\end{equation}%
where in the first terms on the rhs of the previous equations the positive
values of the square roots have been taken.

Let us briefly analyze the physical implications of Eqs.(\ref{acc-1}).

\begin{itemize}
\item The first is that Eqs.(\ref{acc-1}) generate a \emph{diagonal special
NLPT} in which non-local effects are carried only by the time and radial
components of the\textbf{\ }$4-$displacement, i.e., of $4-$velocity and
correspondingly of the $4-$acceleration.

\item The corresponding NLPT which map respectively either the Schwarzschild
(A) or the Reissner-Nordstr\"{o}m (B) space-times onto the Minkowski (C) or
Schwartzchild-analog (D) space-times are provided in all cases by Eqs.(\ref%
{diagonal NLPT}). In particular, the acceleration transformation (\ref%
{acceleration-NLPT}) implies that a point particles endowed with a $4-$%
acceleration $\frac{D^{\prime }}{Ds}u^{\prime \mu }$ with respect to the
space-time C or D, in the space time A or B mapped via Eqs.(\ref{diagonal
NLPT}) is necessarily endowed with an acceleration $\frac{D}{Ds}u^{\mu }$
given by the same equations (i.e., Eqs.(\ref{acceleration-NLPT})).
\end{itemize}

\bigskip

\section{6 - Application \#2: NLPT-acceleration effects in the FLRW
space-time}

Let us now consider the case of the Frieman-Lemaitre-Robertson-Walker (FLRW)
space-time, which incidentally is again of the type (\ref{DIAGONAL CASE}).
Indeed this is obtained identifying respectively%
\begin{equation}
\left\{
\begin{array}{c}
a(r)=1 \\
b(r)=b_{o}^{2}(t) \\
g(r)=b_{o}^{2}(t)\left( \frac{\overline{r}}{r}\right) ^{2}%
\end{array}%
\right.  \label{FLRW-0}
\end{equation}%
where $b_{o}(t)>0$ is a smooth function of the coordinate time, $r$ is the
radial-like coordinate and $\overline{r}$ is prescribed respectively as
\begin{equation}
\overline{r}=\left\{
\begin{array}{ccc}
R_{C}\sinh \left( r/R_{C}\right) &  & \text{for negative curvature }R_{C} \\
r &  & \text{for vanishing curvature }R_{C} \\
R_{C}\sin \left( r/R_{C}\right) &  & \text{for positive curvature }R_{C}%
\end{array}%
\right. ,  \label{FLRW-1}
\end{equation}%
with $R_{C}$ being the curvature associated with the isotropic Ricci tensor
\cite{Ellis}. Let us pose also in this case the problem of the construction
of a mapping onto the Minkowski (C) or the Schwartzchild-analog (D) space
times. In difference with Application \#1 however, here we consider the case
of a non-diagonal special NLPT of the form:%
\begin{equation}
\left\{
\begin{array}{c}
dr^{0}=M_{0}^{0}dr^{\prime 0}+M_{1}^{0}dr^{\prime 1} \\
dr^{1}=M_{0}^{1}dr^{\prime 0}+M_{1}^{1}dr^{1} \\
dr^{i}=M_{(i)}^{i}dr^{\prime (i)} \\
\text{(}i=2,3\text{)}%
\end{array}%
\right. ,
\end{equation}%
with the corresponding inverse transformations given by%
\begin{equation}
\left\{
\begin{array}{c}
dr^{\prime 0}=\frac{dr^{0}}{M_{0}^{0}}-\frac{M_{1}^{0}}{M_{0}^{0}}\frac{%
M_{0}^{0}dr^{1}-M_{0}^{1}dr^{0}}{M_{0}^{0}M_{1}^{1}-M_{0}^{1}M_{1}^{0}} \\
dr^{\prime 1}=\frac{M_{0}^{0}dr^{1}-M_{0}^{1}dr^{0}}{%
M_{0}^{0}M_{1}^{1}-M_{0}^{1}M_{1}^{0}} \\
dr^{\prime (i)}=\frac{dr^{i}}{M_{(g)(i)}^{i}} \\
\text{(}i=2,3\text{)}%
\end{array}%
\right. .
\end{equation}%
Then, in validity of the constraint%
\begin{equation}
S_{0}^{\prime }(r^{\prime })M_{0}^{0}M_{1}^{0}=S_{1}^{\prime }(r^{\prime
})M_{1}^{1}M_{0}^{1},
\end{equation}%
the tensor transformation for the metric tensor yields for case D the roots%
\begin{equation}
\left\{
\begin{array}{c}
M_{0}^{0}=\sqrt{f^{\prime }(r^{\prime })+b_{o}^{2}(t)\left( M_{0}^{1}\right)
^{2}} \\
M_{1}^{1}=\frac{1}{b_{o}(t)}\sqrt{f^{\prime }(r^{\prime
})+b_{o}^{2}(t)\left( M_{0}^{1}\right) ^{2}} \\
M_{1}^{0}=\frac{b_{o}(t)}{f^{\prime }(r^{\prime })}M_{0}^{1} \\
M_{2}^{2}=M_{3}^{3}=\frac{r^{\prime }r}{\overline{r}}%
\end{array}%
\right. ,  \label{roots-case-2}
\end{equation}%
with $M_{0}^{1}$ to be still suitably prescribed. The corresponding
solutions for case C (Minkowski space-time) are obtained letting $f^{\prime
}(r^{\prime })=1$ in the previous equations. Let us briefly analyze the
physical implications of Eqs.(\ref{roots-case-2}):

1) First, we stress that the matrix element $M_{0}^{1}$ is still
undetermined both in magnitude and sign.

2) In this case the time-components of both the $4-$velocity and the $4-$%
acceleration in the FLRW space-time are generated by two "components" acting
in the Minkowski (or Schwartzchild-analog) space-time, respectively the time
and radial components of the $4-$velocity and $4-$acceleration. In
particular one obtains that%
\begin{equation}
\frac{D}{Ds}u^{0}=\sqrt{f^{\prime }(r^{\prime })+b_{o}^{2}(t)\left(
M_{0}^{1}\right) ^{2}}\frac{D^{\prime }}{Ds}u^{\prime 0}+\frac{b_{o}(t)}{%
f^{\prime }(r^{\prime })}M_{0}^{1}\frac{D^{\prime }}{Ds}u^{\prime 1}.
\end{equation}

3) We remark in particular that the second term on the rhs depends in turn
linearly in terms of the arbitrary quantity $M_{0}^{1}$ as well as the
time-dependent factor $b_{o}(t).$ Depending also on the behavior of the
function $b_{o}(t)$ such a contribution may therefore give rise in principle
both to an increase or a decrease in time of the same components.

4) Finally, the radial component of the $4-$acceleration in the FLRW
space-time is generated again by two components in the mapped space-times C
or D, since%
\begin{equation}
\frac{D}{Ds}u^{1}=M_{0}^{1}\frac{D^{\prime }}{Ds}u^{\prime 0}+\frac{1}{%
b_{o}(t)}\sqrt{f^{\prime }(r^{\prime })+b_{o}^{2}(t)\left( M_{0}^{1}\right)
^{2}}\frac{D^{\prime }}{Ds}u^{\prime 1}.
\end{equation}%
Notice that the radial component of the acceleration in the Minkowski or the
Schwartzchild-analog space-time (see, i.e., the 2nd term on the rhs of the
previous equation), gives rise to a corresponding time-dependent
acceleration in the FLRW space-time. The time component contribution on the
rhs is instead proportional to the matrix element $M_{0}^{1}$. As a
consequence also this contribution can in principle significantly affect the
radial acceleration.

\section{7 - Application \#3: NLPT-acceleration effects in Kerr-Newman and
Kerr space-times}

As a further example, let us consider the case of Kerr-Newman and Kerr
space-times, identified here for definiteness with the primed space-time%
\textbf{\ }$\mathbf{(Q}^{\prime 4},g^{\prime }(r^{\prime }))$. In both
cases, when cast in spherical coordinates $(r^{\prime },\theta ^{\prime
},\varphi ^{\prime })$ the corresponding metric tensor is of the generic
non-diagonal form%
\begin{equation}
g_{\mu \nu }^{\prime }(r^{\prime })=\left\vert
\begin{array}{cccc}
S_{0}^{\prime }(r^{\prime }) &  &  & S_{03}^{\prime }(r^{\prime }) \\
& -S_{1}^{\prime }(r^{\prime }) &  &  \\
&  & -S_{2}^{\prime }(r^{\prime }) &  \\
S_{03}^{\prime }(r^{\prime }) &  &  & -S_{3}^{\prime }(r^{\prime })%
\end{array}%
\right\vert .
\end{equation}%
For definiteness, let us first introduce the standard notations%
\begin{equation}
\left\{
\begin{array}{c}
\alpha =\frac{J}{Mc} \\
\rho ^{2}=r^{2}+\alpha ^{2}\cos ^{2}\theta \\
\Delta =r^{2}-rr_{s}+\alpha ^{2}+r_{Q}^{2}%
\end{array}%
\right. ,  \label{notaziuni-Kerr-Newman}
\end{equation}%
where $\alpha $ identifies a constant scale length and $r_{s}$ and $r_{Q}$
the Schwarzschild and Reissner-Nordstr\"{o}m radii (see Eqs.(\ref{radius-1})
and (\ref{radius-2})). Then, the Kerr-Newman metric is defined:%
\begin{equation}
\left\{
\begin{array}{c}
S_{0}^{\prime }(r^{\prime })=\frac{\Delta +\alpha ^{2}\sin ^{2}\theta }{\rho
^{2}} \\
S_{1}^{\prime }(r^{\prime })=\frac{\rho ^{2}}{\Delta } \\
S_{2}^{\prime }(r^{\prime })=\rho ^{2} \\
S_{3}^{\prime }(r^{\prime })=\frac{\Delta }{\rho ^{2}}\alpha ^{2}\sin
^{4}\theta ^{\prime }+\left( r^{\prime 2}+\alpha ^{2}\right) ^{2}\frac{\sin
^{2}\theta ^{\prime }}{\rho ^{2}} \\
S_{03}^{\prime }(r^{\prime })=S_{30}^{\prime }(r^{\prime })=\alpha \left(
r^{\prime 2}+\alpha ^{2}\right) \frac{\sin ^{2}\theta ^{\prime }}{\rho ^{2}}%
\end{array}%
\right. .  \label{KERR-NEWMAN metric}
\end{equation}%
Instead the Kerr metric is prescribed requiring:%
\begin{equation}
\left\{
\begin{array}{c}
S_{0}^{\prime }(r^{\prime })=1-\frac{r_{s}r^{\prime }}{\rho ^{2}} \\
S_{1}^{\prime }(r^{\prime })=\frac{\rho ^{2}}{\Delta } \\
S_{2}^{\prime }(r^{\prime })=\rho ^{2} \\
S_{3}^{\prime }(r^{\prime })=\left( r^{\prime 2}+\alpha ^{2}+\frac{%
r_{s}r^{\prime }\alpha ^{2}}{\rho ^{2}}-\frac{r_{s}r^{\prime }}{\rho ^{2}}%
\sin ^{2}\theta ^{\prime }\right) \sin ^{2}\theta ^{\prime } \\
S_{03}^{\prime }(r^{\prime })=S_{30}^{\prime }(r^{\prime })=-\frac{%
r_{s}r^{\prime }\alpha }{\rho ^{2}}\sin ^{2}\theta ^{\prime }%
\end{array}%
\right. .
\end{equation}%
Let us now pose the problem of mapping either the Kerr-Newman or the Kerr
space-times onto the Minkowski space-time\textbf{\ }$(\mathbf{M}^{4}.\eta )$%
. For convenience let us represent also the latter space-time in spherical
coordinates. This gives for the Minkowski metric the customary diagonal
representation%
\begin{equation}
\eta _{\mu \nu }(r)\equiv diag\left(
S_{0}(r)=1,-S_{1}(r)=-1,-S_{2}(r)=-r^{2},-S_{3}(r)=-r^{2}\sin ^{2}\theta
\right) .  \label{CWARTZ-1}
\end{equation}%
A possible non-unique realization of the NLPT between the two space-times $%
\mathbf{(Q}^{\prime 4},g^{\prime })$ and $(\mathbf{M}^{4}.\eta )$ indicated
above,\ in some sense analogous to the one developed here for the FLRW
space-time (see Section 6), is proposed here. This is provided by the \emph{%
non-diagonal special NLPT} of the form%
\begin{equation}
\left\{
\begin{array}{c}
dr^{0}=M_{0}^{0}dr^{\prime 0}+M_{1}^{0}dr^{\prime 1}+M_{3}^{0}dr^{\prime 3}
\\
dr^{1}=M_{1}^{1}dr^{\prime 1}+M_{0}^{1}dr^{\prime 0} \\
dr^{2}=M_{2}^{2}dr^{\prime 2} \\
dr^{3}=M_{1}^{3}dr^{\prime 1}+M_{3}^{3}dr^{\prime 3}%
\end{array}%
\right. ,  \label{po-1}
\end{equation}%
subject to the validity of the constraints%
\begin{equation}
\left\{
\begin{array}{c}
S_{0}(r)M_{0}^{0}M_{1}^{0}-S_{1}(r)M_{1}^{1}M_{0}^{1}=0 \\
S_{0}(r)M_{1}^{0}M_{3}^{0}-S_{3}(r)M_{3}^{3}M_{0}^{3}=0%
\end{array}%
\right. .  \label{kerr-constraints}
\end{equation}

One can readily show that Eqs.(\ref{po-1}) indeed realize a NLPT which
mutually maps in each other the two space-times $\mathbf{(Q}^{\prime
4},g^{\prime })$ and $(\mathbf{M}^{4}.\eta )$. For this purpose, in validity
of Eqs.(\ref{CWARTZ-1}), let us require that the Jacobian matrix $M_{\nu
}^{\mu }$ satisfies the further tensor equations%
\begin{equation}
\left\{
\begin{array}{c}
\left( M_{0}^{0}\right) ^{2}-\left( M_{0}^{1}\right) ^{2}=S_{0}^{\prime
}(r^{\prime }) \\
\left( M_{3}^{0}\right) ^{2}-\left( M_{1}^{1}\right) ^{2}-r^{2}\sin
^{2}\theta \left( M_{1}^{3}\right) ^{2}=-S_{1}^{\prime }(r^{\prime }) \\
r^{2}\left( M_{2}^{2}\right) ^{2}=S_{2}^{\prime }(r^{\prime }) \\
\left( M_{1}^{0}\right) ^{2}-r^{2}\sin ^{2}\theta \left( M_{3}^{3}\right)
^{2}=-S_{3}^{\prime }(r^{\prime })%
\end{array}%
\right. .  \label{tensoir equations}
\end{equation}%
From Eqs.(\ref{tensoir equations}) and (\ref{kerr-constraints}) elementary
algebra gives the general solution%
\begin{equation}
\left\{
\begin{array}{c}
M_{1}^{0}=\frac{M_{1}^{1}}{M_{0}^{0}}M_{0}^{1} \\
M_{3}^{0}=\frac{r^{2}\sin ^{2}\theta M_{0}^{0}M_{3}^{3}}{M_{1}^{1}M_{0}^{1}}%
M_{0}^{3} \\
M_{0}^{0}=\sqrt{S_{0}^{\prime }(r^{\prime })+\left( M_{0}^{1}\right) ^{2}}
\\
M_{1}^{1}=\sqrt{\frac{S_{1}^{\prime }(r^{\prime })-r^{2}\left(
M_{1}^{3}\right) ^{2}}{S_{0}^{\prime }(r^{\prime })}}\sqrt{S_{0}^{\prime
}(r^{\prime })+\left( M_{0}^{1}\right) ^{2}} \\
M_{3}^{3}=\sqrt{S_{0}^{\prime }(r^{\prime })+\left( M_{0}^{1}\right) ^{2}}%
\end{array}%
\right. ,  \label{NLPT-KERR solution}
\end{equation}%
where the matrix elements $M_{0}^{1}$ and $M_{1}^{3}$ still remain in
principle arbitrary. Notice that the ratios $\frac{S_{1}(r)M_{1}^{1}}{%
S_{0}(r)M_{0}^{0}}$ and $\frac{S_{3}(r)M_{0}^{0}}{S_{1}(r)M_{1}^{1}}$read
then%
\begin{eqnarray}
\frac{M_{1}^{1}}{M_{0}^{0}} &=&\sqrt{\frac{S_{1}^{\prime }(r^{\prime
})-r^{2}\left( M_{1}^{3}\right) ^{2}}{S_{0}^{\prime }(r^{\prime })}}, \\
\frac{r^{2}\sin ^{2}\theta M_{0}^{0}}{M_{1}^{1}} &=&r^{2}\sin ^{2}\theta
\sqrt{\frac{S_{1}^{\prime }(r^{\prime })-r^{2}\left( M_{1}^{3}\right) ^{2}}{%
S_{0}^{\prime }(r^{\prime })}}.
\end{eqnarray}%
Hence it follows that%
\begin{equation}
\begin{array}{c}
M_{1}^{0}=\sqrt{\frac{S_{1}^{\prime }(r^{\prime })-r^{2}\left(
M_{1}^{3}\right) ^{2}}{S_{0}^{\prime }(r^{\prime })}}M_{0}^{1}, \\
M_{3}^{0}=r^{2}\sin ^{2}\theta \sqrt{\frac{S_{1}^{\prime }(r^{\prime
})-r^{2}\left( M_{1}^{3}\right) ^{2}}{S_{0}^{\prime }(r^{\prime })}}\frac{%
M_{3}^{3}}{M_{0}^{1}}M_{1}^{3}.%
\end{array}%
\end{equation}%
Notice, however, that the existence of the solution (\ref{NLPT-KERR solution}%
) demands manifestly%
\begin{equation}
S_{1}^{\prime }(r^{\prime }).-r^{2}\left( M_{1}^{3}\right) ^{2}>0,
\label{solubility conditiuon}
\end{equation}%
to be interpreted as solubility condition. A number of remarks can be made:

1) Notice that when letting in particular $M_{1}^{0}=M_{1}^{3}$, Eqs.(\ref%
{po-1}) reduce to the non-diagonal special NLPT considered in Part 2.

2) Eqs.(\ref{po-1}) imply that the time-component of the $4-$acceleration in
the Minkowski space-time is generated by time-, radial and tangential
components in the Kerr-Newman and Kerr space-times respectively, namely $%
\frac{D^{\prime }}{Ds}u^{\prime 0},$ $\frac{D^{\prime }}{Ds}u^{\prime 1}$
and $\frac{D^{\prime }}{Ds}u^{\prime 3}$.

3) Similarly, the tangential component $\frac{D}{Ds}u^{3}$ depends also on
the radial component $\frac{D^{\prime }}{Ds}u^{\prime 1}$ arising in the
Kerr-Newman or Kerr space-times, besides $\frac{D^{\prime }}{Ds}u^{\prime 3}$%
.

4) Let us now consider the inverse transformations following from Eqs.((\ref%
{po-1}). In analogy with Application \#2 (see Section 5) also in this case
both the time and radial components arising in the Kerr-Newman or Kerr
space-times, namely respectively $\frac{D^{\prime }}{Ds}u^{\prime 0}$ and $%
\frac{D^{\prime }}{Ds}u^{\prime 1}$ generally depend on the analogous
components of the $4-$acceleration arising in the Minkowski space-time,
namely $\frac{D}{Ds}u^{0}$ and $\frac{D}{Ds}u^{1}$, as well as the
tangential component $\frac{D}{Ds}u^{3}.$ Thus, for example one obtains that%
\begin{equation}
\frac{D^{\prime }}{Ds}u^{\prime 0}=\frac{\left[ M_{3}^{3}M_{1}^{1}\frac{D}{Ds%
}u^{0}+\left( M_{3}^{0}M_{1}^{3}-M_{3}^{3}M_{1}^{0}\right) \frac{D}{Ds}%
u^{1}-M_{1}^{1}M_{3}^{0}\frac{D}{Ds}u^{3}\right] }{\left(
M_{0}^{0}M_{1}^{1}-M_{1}^{0}M_{0}^{1}\right)
M_{3}^{3}-M_{3}^{0}M_{1}^{3}M_{0}^{1}},  \label{time-acceleration}
\end{equation}%
and similarly the radial component reads%
\begin{equation}
\frac{D^{\prime }}{Ds}u^{\prime 1}==\frac{1}{M_{1}^{1}}\left[ \frac{D}{Ds}%
u^{1}-M_{0}^{1}\frac{D^{\prime }}{Ds}u^{\prime 0}\right] .
\label{radiual acceleration}
\end{equation}

5) In the previous equations the matrix elements $M_{0}^{1}$ and $M_{1}^{3}$
remain still in principle arbitrary, with the second one required to fulfill
the inequality (\ref{solubility conditiuon}) indicated above. A further
solubility condition is provided, however, by\ the equation for $M_{3}^{0}$
in Eqs.(\ref{NLPT-KERR solution}). In fact, this is only defined provided
also%
\begin{equation}
M_{0}^{1}\neq 0.  \label{solubility-2}
\end{equation}

6) Since due to Eqs.(\ref{kerr-constraints}) the matrix elements $M_{1}^{0}$
and $M_{3}^{0}$ become linear functions of $M_{0}^{1}$ and $M_{1}^{3}$
respectively, it follows that both the time and radial accelerations (\ref%
{time-acceleration}) and (\ref{radiual acceleration}) strongly depend on the
choices of the same parameters. Notice that, in particular, the coupling of $%
\frac{D^{\prime }}{Ds}u^{\prime 1}$ with non-radial components occurs \emph{%
always} due to the solubility condition (\ref{solubility-2}).

\section{8 - Application \#4: NLPT-transformation laws of the EM Faraday
tensor}

Let us now consider as an application of NLPT-theory\ the dynamics of a
charged point particle of rest-mass $m_{o}$ and electric charge $q$ immersed
in an external EM field. As it is well known in the curved space-times $(%
\mathbf{Q}^{4},g)$ and $(\mathbf{Q}^{\prime 4},g^{\prime })$ the
relativistic equation of motion takes respectively the forms%
\begin{eqnarray}
m_{o}\frac{Du^{\mu }}{Ds} &=&qF_{\nu }^{(ext)\mu }(r)u^{\nu }, \\
m_{o}\frac{D^{\prime }u^{\prime \mu }}{Ds} &=&qF_{\nu }^{\prime (ext)\mu
}(r^{\prime })u^{\prime \nu }
\end{eqnarray}%
with $u^{\mu },$ $u^{\prime \mu }$ and respectively $F_{\nu }^{(ext)\mu
}(r), $ $F_{\nu }^{\prime (ext)\mu }(r^{\prime })$ denoting the in the same
space-times the $4-$velocities and the Faraday tensors generated by an
externally-produced EM field. Assuming that a general NLPT maps in each
other $(\mathbf{Q}^{4},g)$ and $(\mathbf{Q}^{\prime 4},g^{\prime })$ the
NLPT-transformation law for the $4-$acceleration, and in particular Eqs.(\ref%
{T3-3}) and (\ref{T3-4}) in THM.2, since%
\begin{equation*}
\frac{D}{Ds}u^{\mu }=M_{(g)\nu }^{\mu }(r^{\prime },r)\frac{D^{\prime }}{Ds}%
u^{\prime \nu }
\end{equation*}%
it must be identically that%
\begin{equation}
qF_{\alpha }^{(ext)\mu }(r)u^{\alpha }=M_{(g)\beta }^{\mu }(r^{\prime
},r)qF_{\nu }^{\prime (ext)\beta }\left( M^{-1}\right) _{(g)\alpha }^{\nu
}(r,r^{\prime })u^{\prime \alpha }.
\end{equation}%
Therefore, due to the arbitrariness of the $4-$vector $u^{\prime \alpha },$
the quantity $F_{\alpha }^{(ext)\mu }(r)$ necessarily satisfies the $4-$%
tensor NLPT (direct) transformation law
\begin{equation}
F_{\alpha }^{(ext)\mu }(r)=M_{(g)\beta }^{\mu }(r^{\prime },r)F_{\nu
}^{\prime (ext)\beta }(r^{\prime })\left( M^{-1}\right) _{(g)\alpha }^{\nu
}(r,r^{\prime })  \label{direct NLPT-transformation lwa for F}
\end{equation}%
Hence by construction it follows%
\begin{equation*}
\left( M^{-1}\right) _{(g)\mu }^{k}F_{\alpha }^{(ext)\mu }M_{(g)j}^{\alpha
}=\left( M^{-1}\right) _{(g)\mu }^{k}M_{(g)\beta }^{\mu }F_{\nu }^{\prime
(ext)\beta }\left( M^{-1}\right) _{(g)\alpha }^{\nu }M_{(g)j}^{\alpha }
\end{equation*}%
which yields the corresponding inverse transformation law too%
\begin{equation}
F_{j}^{\prime (ext)k}(r^{\prime })=\left( M^{-1}\right) _{(g)\mu
}^{k}(r,r^{\prime })F_{\alpha }^{(ext)\mu }(r)M_{(g)j}^{\alpha }(r^{\prime
},r).  \label{inverse transf.law for F}
\end{equation}%
Eqs.(\ref{direct NLPT-transformation lwa for F}) and (\ref{inverse
transf.law for F}) provide the transformation equations connecting for the
Faraday tensors $F_{\alpha }^{(ext)\mu }(r)$ and $F_{\alpha }^{\prime
(ext)\mu }(r^{\prime })$ which are defined respectively in the two
space-times $(\mathbf{Q}^{4},g)$ and $(\mathbf{Q}^{\prime 4},g^{\prime }).$
In particular we stress that on the rhs of the first equation $r\equiv
\left\{ r^{\mu }\right\} $ must be regarded as a non-local function of $%
r^{\prime }\equiv \left\{ r^{\prime \mu }\right\} $ whose form is determined
by the same NLPT. This means that $F_{\nu }^{(ext)\mu }(r)$ (and conversely $%
F_{\alpha }^{\prime (ext)\beta }(r^{\prime })$ when represented via the
inverse transformation (\ref{inverse transf.law for F})) must be regarded in
turn as a non-local function of $r^{\prime }$ too$.$

There remains an important question to answer, i.e., whether the transformed
Faraday tensor $F_{\alpha }^{\prime (ext)\beta }$ can be identified or not
with an exact solution of the Maxwell equations defined in the space-time $(%
\mathbf{Q}^{4},g)$ $,$ being $F_{\nu }^{(ext)\mu }(r)$ an exact solution of
the same equations in $(\mathbf{Q}^{\prime 4},g^{\prime })$. The answer to
this question requires to prove that $F_{\nu }^{(ext)\mu }(r)$ and $F_{\nu
}^{\prime (ext)\mu }(r^{\prime })$ are respectively solutions of the Maxwell
equations in the two space-times, i.e., that these equations are endowed
with a tensor transformation law with respect to the group of general NLPT $%
\left\{ P_{g}\right\} .$ The proof of this statement will be reported
elsewhere.

\section{9 - Physical implications and conclusions}

In this paper implications of NLPT-theory presented in Parts 1 and 2 have
been investigated with particular reference to the establishment of the $4-$%
acceleration tensor transformations. In our view the analysis here performed
can help reaching a novel interpretation and a deeper physical
interpretation of the standard formulation of GR (SF-GR).

The NLPT-theory here developed is based on\ the extension of\ the customary
functional setting which lays at the\ basis of the same SF-GR. This involves
the construction of a suitable invertible coordinate transformation, to be
identified with a suitable class of non-local point transformation (NLPT)
which map in each other two in principle arbitrary curved space-times $(%
\mathbf{Q}^{4},g)$ and $(\mathbf{Q}^{\prime 4},g^{\prime })$ possibly
parametrized in terms of different curvilinear coordinate systems. In other
words, the two transformed space-times may exhibit, as a consequence,
intrinsically-different non-vanishing Riemann curvature tensors. As pointed
out in Part 1 the new transformations have the distinctive property that the
corresponding Jacobians are endowed with a characteristic non-gradient form.
\ In particular these transformation, identified with the general NLPT-group
$\left\{ P_{g}\right\} ,$ involve also to the adoption of a new kind of
phase-space reference frames, denoted as extended GR-frames involving the
prescription of both $4-$position and corresponding $4-$velocities (see
related discussion in Part 1).

In this paper in particular the fundamental issue has been addressed of the
establishment of the tensor transformation laws relating the $4-$%
acceleration arising in different space-times as well as of the
investigation of the main related physical implications.

In order to reach the goal indicated above in this paper a systematic
analysis of the mathematical implications of NLPT-theory has been
performed.\ This has been realized, first, by inspecting the differential
properties of the Jacobian of general NLTP, established by Lemmas 1 and 2
(see Section 2). Second, the transformation properties of the Christoffel
symbols with respect to the NLPT-group $\left\{ P_{g}\right\} $ have been
determined (see THM.1 in Section 3). Notably, this includes also as a
particular possible realization in the case of NLPT mapping a curved
space-time onto the Minkowski space-time (Corollary to THM.1). Based on
these results, the tensor transformation laws of the $4-$accelerations
within the general NLPT-group $\left\{ P_{g}\right\} $ have been established
(see THM.2, Section 4), thus enabling one to determine the relationships
holding between the $4-$accelerations which are defined in the two different
curved space-times $\left\{ \mathbf{Q}^{4},g\right\} $ and $\left\{ \mathbf{Q%
}^{\prime 4},g^{\prime }\right\} $ mutually mapped in each other by an
arbitrary NLPT.

Physical insight about the meaning and implications of the tensor
transformation laws determined here emerges from the applications considered
in Sections 5, 6, 7 which have concerned the analysis of NLPT-acceleration
effects arising:

\begin{itemize}
\item Between the Schwarzschild, Reissner-Nordstr\"{o}m\emph{\ }and the
Minkowski or Schwarzschild-analog space-times.

\item Between the Friedman-Lemaitre-Robertson-Walker and the Minkowski or
Schwarzschild-analog space-times.

\item Between the Kerr-Newman or Kerr space-times and the Minkowski
space-time.
\end{itemize}

as well as, notably, Section 8 dealing with:

\begin{itemize}
\item the determination of the NLPT-tensor transformation laws of the EM
Faraday tensor, a result which is directly.implied by the $4-$acceleration
transformation law here established.
\end{itemize}

These results are undoubtedly in qualitative consistency with the Einstein
famous equivalence principle (EEP, \cite{Einst}; see also related discussion
in Part 1)) and, more precisely with Einstein's key related conjecture which
actually lays at the basis of GR, namely that \textquotedblleft \emph{local
effects of motion in a curved space} (\emph{produced by gravitation}%
)\textquotedblright\ should be considered\ as\ \textquotedblleft \emph{%
indistinguishable from those of an accelerated observer in flat space}%
\textquotedblright\ \cite{ein-1907,ein-1911}.

In order to further elucidate the issue let us in fact consider a NLPT
transforming in each other a curved space-time $\left( \mathbf{Q}%
^{4},g\right) $ and the flat Minkowski space-time \ $\left( \mathbf{M}%
^{4},\eta \right) .$ In addition, for definiteness let us require that in
both cases the same coordinate systems are adopted (so that the NLPT
identifies a special NLPT). Then denoting $a^{\mu }$ and $a^{\prime \mu }$
the accelerations in the two space-times -  in view of the discuessions of
the aapplications \#1-\#3 discussed above - it is always possible to
construct the NLPT in such a way that for example the  equation
\begin{equation}
0\equiv a^{1}=M_{0}^{1}a^{\prime 0}+M_{1}^{1}a^{\prime 1}
\end{equation}%
holds identically, i.e., the spatial component of the acceleration in the
curved space-time ($a^{1}$) vanishes identically, while the corresponding
component in the Minkowski space-time is non-vanishing. In other words, the
acceleration effect arising in the flat space-time apparently disappears in
the curved space-time (i.e., is effectively hiddent in the
gravitational-curvature effects of the space-time $\left( \mathbf{Q}%
^{4},g\right) .$

To conclude, it should also be mentioned that also the transformation law of
the EM Faraday tensor and the NLPT-covariance law of Maxwell equations here
discovered are actually consistent with the same Einstein's viewpoint. In
fact, in accordance with Einstein, \emph{exclusively the effects on particle
motion which are due to gravitation should be considered indistinguishable
from those of an accelerated observer}. \ \

As discussed at length also in Parts 1 and 2 of the present investigation,
both accelerations effects are found to be realized in the framework of
NLPT-theory. In turn, this requires the adoption of phase-space reference
frames, denoted as extended GR-frames, and a suitably-defined set of
phase-space maps, which involve in particular the introduction of
appropriate non-local coordinate transformations identified with the group
of general NLPT $\left\{ P_{g}\right\} $\emph{.}

These conclusions strongly support the crucial importance of non-locality
effects in physics. In particular, in our view, the NLPT-theory here
presented appears susceptible of a plethora of potential applications from
classical relativistic mechanics and electrodynamics \cite%
{EPJ1,EPJ2,EPJ3,EPJ4,EPJ5,EPJ6}, general relativity quantum theory of
extended particle dynamics \cite{dewitt,quinn,crowley,wald,EPJ7},
relativistic kinetic theory \cite{EPJ5}, to cosmology as well as
relativistic quantum mechanics \cite{EPJ8} and quantum gravity.\textbf{\ }%
\bigskip

\section{Acknowledgments}

Work developed within the research projects of the Czech Science Foundation
GA\v{C}R grant No. 14-07753P (C.C.)\ and Albert Einstein Center for
Gravitation and Astrophysics, Czech Science Foundation No. 14-37086G (M.T.).


\begin{thebibliography}{99}
\bibitem{noi1} M. Tessarotto and C. Cremaschini, \textit{Theory of non-local
point transformations - Part 1: \ Representation of Teleparallel Gravity},
Eur. Phys. J. Plus, submitted (2015).

\bibitem{noi2} M. Tessarotto and C. Cremaschini, \textit{Theory of non-local
point transformations - Part 2: General form and Gedanken experiment}, Eur.
Phys. J. Plus, submitted (2015).

\bibitem{Einstein1915} A. Einstein, \textit{Die Feldgleichungen der
Gravitation, }Sitzungsber. Preuss. Akad. Wiss. (Berlin), 844 (1915).

\bibitem{Einst} A. Einstein, \textit{The Meaning of Relativity,} Princeton
University Press (1945).

\bibitem{Landau} L.D. Landau and E.M. Lifschitz, \textit{The classical
theory of fields, Vol.2} (Addison-Wesley, N.Y., 1957).

\bibitem{Wheeler} J.A. Wheeler, C. Misner, K.S. Thorne, \textit{Gravitation,}
W.H. Freeman \& Co (1973).

\bibitem{Wheeler2} ibid., pp. 85--86, \S 3.5 (1973).

\bibitem{Synge} J.L. Synge, A. Schild, \textit{Tensor Calculus. first Dover
Publications} 1978 edition. pp. 6--108 (1949).

\bibitem{ein-1907} A. Einstein, \textit{Relativit\"{a}tsprinzip und die aus
demselben gezogenen Folgerungen} (\textit{On the Relativity Principle and
the Conclusions Drawn from It}). Jahrbuch der Radioaktivit\"{a}t \textbf{4},
411 (1907).

\bibitem{ein-1911} A. Einstein, \textit{Einfluss der Schwerkraft auf die
Ausbreitung des Lichtes} (\textit{On the Influence of Gravitation on the
Propagation of Light}), Annalen der Physik \textbf{35}, 898 (1911).

\bibitem{Einstein1928} A. Einstein, Preuess. Akad. Wiss., 414 (1925);
Sitzungsber. Preuss. Akad. Wiss. Phys. Math. Kl. 217 (1928); 401 (1930);
Math. Ann. 102 , 685 (1930).

\bibitem{mash1} U. Muench, F.W. Hehl, B. Mashhoon, Phys. Lett. A \textbf{271}%
, 8 (2000).

\bibitem{mash2} B. Mashhoon, Annalen der Physik \textbf{523}, 226 (2011).

\bibitem{tele1} P. Wu and H. Yu, Phys. Lett. B \textbf{693}, 415 (2010).

\bibitem{tele2} K. Bamba, C.Q. Geng, C.C. Lee and L.-W. Luo, J. Cosmol.
Astropart. Phys. \textbf{01}, 021 (2011).

\bibitem{tele3} T.P. Sotiriou, B. Li and J.D. Barrow, Phys. Rev. D \textbf{83%
},104030 (2011).

\bibitem{NJ1} E.T. Newman and A.I. Janis, J. Math. Phys. \textbf{6}, 915
(1965).

\bibitem{NJ2} E.T. Newman, E. Couch, K. Chinnapared, A. Exton, A. Prakash
and R. Torrence, J. Math. Phys. \textbf{6}, 918 (1965).

\bibitem{NJ3} S.P. Drake and P. Szekeres, Gen. Relativ. Gravit. \textbf{32},
445 (2000).

\bibitem{bambi} C. Bambi and L. Modesto, Phys. Lett. B \textbf{721}, 329
(2013).

\bibitem{bambi2} B. Toshmatov, B. Ahmedov, A. Abdujabbarov and Z. Stuchl%
\'{\i}k, Phy

\bibitem{Cartan1922} E. Cartan, C. R. Acad. Sci. Paris 174, 593 (1922); 174,
734 (1922).

\bibitem{Ellis} G. F. R. Ellis, in \textit{Vth Brazilian School on Cosmology
and Gravitation}, Ed. M Novello (World Scientific, Singapore, 1987), 83.

\bibitem{EPJ1} C. Cremaschini and M. Tessarotto, Eur. Phys. J. Plus \textbf{%
126}, 42 (2011).

\bibitem{EPJ2} C. Cremaschini and M. Tessarotto, Eur. Phys. J. Plus \textbf{%
126}, 63 (2011).

\bibitem{EPJ3} C. Cremaschini and M. Tessarotto, Eur. Phys. J. Plus \textbf{%
127}, 4 (2012).

\bibitem{EPJ4} C. Cremaschini and M. Tessarotto, Eur. Phys. J. Plus \textbf{%
127}, 103\textbf{\ }(2012).

\bibitem{EPJ5} C. Cremaschini and M. Tessarotto, Phys. Rev. E \textbf{87},
032107 (2013).

\bibitem{EPJ6} C. Cremaschini and M. Tessarotto, Int. J. Mod. Phys. A
\textbf{28}, 1350086 (2013).

\bibitem{EPJ7} C. Cremaschini and M. Tessarotto, Eur. Phys. J. Plus \textbf{%
129}, 247 (2014).

\bibitem{EPJ8} C. Cremaschini and M. Tessarotto, Eur. Phys. J. Plus \textbf{%
130}, 166 (2015).

\bibitem{dewitt} B.S. DeWitt and R.W. Brehme, Ann. Phys. \textbf{9}, 220
(1960).

\bibitem{crowley} R.J. Crowley and S.J. Nodvik , Ann. Phys. \textbf{113}, 98
(1978).

\bibitem{wald} T.C. Quinn and R.M. Wald, Phys. Rev. D \textbf{56}, 3381
(1997).

\bibitem{quinn} T.C. Quinn, Phys. Rev. D \textbf{62}, 064029 (2000).

\bibitem{q1} E.J. Moniz and D.H. Sharp, Phys. Rev. D \textbf{15}, 2850
(1977).

\bibitem{q4} P.R. Johnson and B.L. Hu, Phys. Rev. D \textbf{65}, 065015
(2002).

\bibitem{q3} A. Higuchi and G.D.R. Martin, Found. Phys. \textbf{35}, 7
(2005).
\end{thebibliography}
\end{document}